\documentclass[11pt]{amsart}
\usepackage{lmodern}
\usepackage[T1]{fontenc}
\usepackage{microtype}
\usepackage{amssymb}
\usepackage{amsthm}
\usepackage{amscd}
\usepackage{hyperref}
\usepackage{cite}
\usepackage{color}

\usepackage{amsmath}
\usepackage{tikz}
\usetikzlibrary{matrix,arrows}
\usepackage{amsfonts,amssymb,amsthm, txfonts, pxfonts,amscd}

\usepackage{soul,color}

\def\pr{\mathop{\text{pr}}\nolimits}

\makeatletter

\def\dotminussym#1#2{%
  \setbox0=\hbox{$\m@th#1-$}%
  \kern.5\wd0%
  \hbox to 0pt{\hss\hbox{$\m@th#1-$}\hss}%
  \raise.6\ht0\hbox to 0pt{\hss$\m@th#1.$\hss}%
  \kern.5\wd0}

\mathchardef\mhyphen="2D

\begin{document}

\title{Atypical scaling behavior persists in real world interaction networks} 
\author{Harry Crane and Walter Dempsey}
\address {Department of Statistics \& Biostatistics, Rutgers University, 110 Frelinghuysen Avenue, Piscataway, NJ 08854, USA}
\email{hcrane@stat.rutgers.edu}
\urladdr{\url{http://stat.rutgers.edu/home/hcrane}}
\address {Department of Statistics, University of Michigan, 1085 S. University Ave,  Ann Arbor, MI 48109, USA}
\email{wdem@umich.edu}
\thanks{H.\ Crane is partially supported by NSF grant DMS-1308899 and NSA grant H98230-13-1-0299.}

\date{July 13, 2015}

\begin{abstract}

Scale-free power law structure describes complex networks derived from a wide range of real world processes.
The extensive literature focuses almost exclusively on networks with power law exponent strictly larger than 2, which can be explained by constant vertex growth and  preferential attachment.
The complementary scale-free behavior in the range between 1 and 2 has been mostly neglected as atypical because there is no known generating mechanism to explain how networks with this property form.
However, empirical observations reveal that scaling in this range is an inherent feature of real world networks obtained from repeated interactions within a population, as in social, communication, and collaboration networks.
A generative model explains the observed phenomenon through the realistic dynamics of constant edge growth and a positive feedback mechanism.
Our investigation, therefore, yields a novel empirical observation grounded in a strong theoretical basis for its occurrence.

 \end{abstract}

\maketitle

\begin{figure}
\includegraphics[scale=0.8]{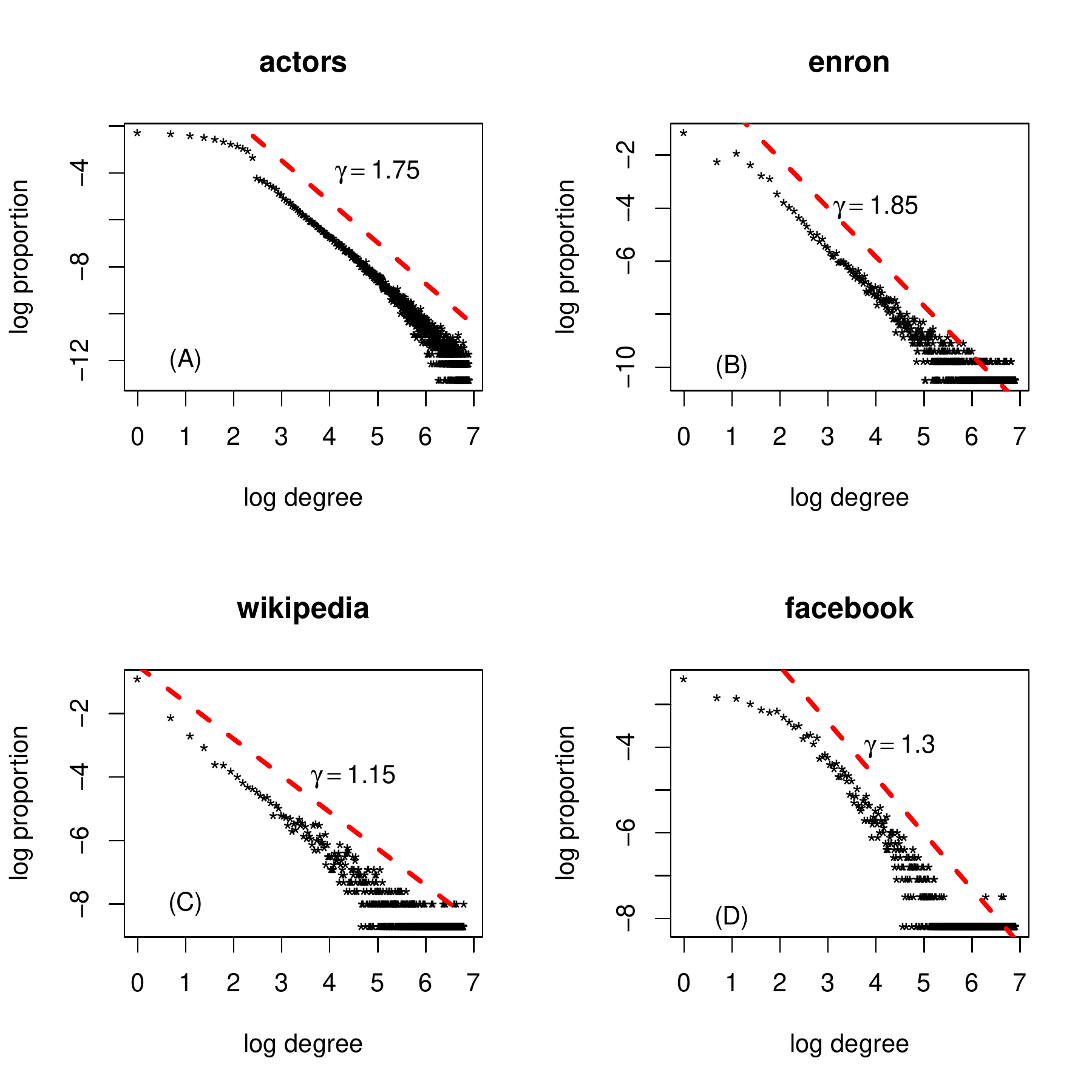}
\caption{Empirical degree distributions for (A) Actors collaboration network, (B) Enron e-mail network, (C) Wikipedia link network, and (D) Facebook social circles network.  In each panel, the slope of the dashed line is $-\gamma$, where $\gamma$ is the estimated power law exponent.
Fitting the two-parameter generative model in \eqref{eq:update} to the data, we obtain estimates (A) $(\alpha_{\text{actor}},\theta_{\text{actor}})=(0.75,1.14)$, (B) $(\alpha_{\text{enron}},\theta_{\text{enron}})=(0.85,-0.63)$, (C) $(\alpha_{\text{wiki}},\theta_{\text{wiki}})=(0.15,350)$, and (D) $(\alpha_{\text{fb}},\theta_{\text{fb}})=(0.30,168)$.}
\label{fig:alldatasets}
\end{figure}

Self-organizing dynamics of many processes produce a common heterogeneous structure characterized by power law degree distributions, which have been discovered in the World Wide Web \cite{BA1999,FFF1999,Kumar1999}, social networks \cite{WattsStrogatz1998}, telecommunications networks \cite{Abello1998}, biological networks \cite{JeongMason2001}, and many others \cite{ChungLubook,DorogovtsevMendes2003,Durrett2010PNAS,Newman2003}.
A network exhibits {\em power law} degree distribution with exponent $\gamma>1$ if the proportion $p_k$ of vertices with degree $k$ satisfies $p_k\sim k^{-\gamma}$ for large $k$.
Figure \ref{fig:alldatasets} plots the degree distributions of four well known networks: the actors collaboration network \cite{BA1999}, Enron email network \cite{McAuleyLeskovec2012}, Wikipedia voting network \cite{LeskovecHuttenlocherKleinber2010}, and Facebook social circles network \cite{KlimtYang2004,LeskovecLangDasguptaMahoney2009}.
The power law exponent in each of these networks is between 1 and 2, behavior that cannot be explained by preferential attachment models.
Preferential attachment dynamics provide an intuitive description of networks that undergo constant vertex growth and exhibit power law greater than 2.
These properties do not accurately describe the networks in Figure \ref{fig:alldatasets}:
\begin{itemize}
	\item[(A)] In the actor collaboration network, vertices correspond to actors and edges represent that two actors were cast in the same movie.
Here the data permits multiple edges between vertices if the actors were cast together more than once.
Thus, the network grows as a consequence of movie production, i.e., edge formation, rather than the influx of new actors.
	\item[(B)] In the Enron email network, vertices correspond to employees at the Enron corporation and an edge represents that an email has been exchanged between those employees.
As emails are exchanged, new edges form without any requirement that new vertices be added.
	\item[(C)] The Wikipedia voting network represents voting behavior for elections to the administrator role in Wikipedia.
Vertices are Wikipedia users and a directed edge points from $i$ to $j$ if user $i$ voted for user $j$.
The network grows when elections are held, i.e., new edges are formed.
	\item[(D)] The Facebook social circles network represents friendships among Facebook users.
The network grows by the formation of new friendships, i.e., edges, which usually result from social interactions among users. 
\end{itemize}
Each of the above networks grows by the addition of edges that connect according to a positive feedback mechanism, whereby past interactions reinforce future behavior.
For example, an email sent from employee A to employee B is likely to be reciprocated by a reply from B to A; actors cast together in one movie likely play complementary roles which may be suitable in future movies; and so on. 

While positive feedback exhibits obvious similarities to preferential attachment, it differs in that edge formation need not be accompanied by the addition of new vertices.
Furthermore, the range of the power law exponent implies additional growth properties about the network.
Power law exponent $\gamma>2$ implies that the expected vertex degree grows at rate $\sum_{k=1}^nk\cdot k^{-\gamma}\approx\int_{1}^n x^{-\gamma+1}dx\sim O(1)$ as a function of the number of vertices, making the total number of edges grow at the rate $n\cdot O(1)=O(n)$.
Therefore, preferential attachment models implicitly assume a network for which the number of edges grows linearly with the number of vertices.
On the other hand, for $1<\gamma<2$, the expected degree grows at rate $\sum_{k=1}^{n}k\cdot k^{-\gamma}\sim O(n^{2-\gamma})$ as a function of the number of vertices $n$, indicating total edge growth at rate $n\cdot O(n^{2-\gamma})=O(n^{3-\gamma})$ in the intermediate range between sparsity $O(n)$ and density $O(n^2)$.

Some recent progress in the mathematical literature demonstrates the fundamentally different structural properties of sparse and dense networks \cite{BorgsChayes2014Lp1,BorgsChayes2014Lp2,LovaszSzegedy2006}.
Figure \ref{fig:alldatasets} suggests that power law exponent between 1 and 2 is also of scientific interest, and understanding this intermediate range should provide important insights into the structure of real world networks. 
We replicate these features in the following generative model, which produces scale-free networks with exponent $1<\gamma<2$ and closely resembles how networks (A)-(D) form.
We generate a network with $n$ edges by sequentially adding one edge at each time $t=1,2,\ldots,n$.
Our model is determined by two parameters $\alpha$ and $\theta$ in the range $0<\alpha<1$ and $\theta>-\alpha$.
Before time $t$, the network has $t-1$ edges and a random number of vertices $N_t$, with the initial condition $N_1=0$.
We label these vertices $i=1,\ldots,N_t$ and write $D(i,t)$ to denote the total degree of vertex $i$ before the $t$th edge is added.
(Note that each self-loop from a vertex to itself contributes 2 to its degree.)
When the $t$th edge arrives, its two incident vertices $v_1(t),v_2(t)$ are chosen randomly among vertices $1,\ldots,N_t$ and a new vertex $N_t+1$ as follows.
With $N_t^1=N_t$, we first choose $v_1(t)$ randomly with probability
\begin{equation}\label{eq:update}
\pr(v_1(t)=i)\propto\left\{\begin{array}{cc}
D(i,t)-\alpha,& i=1,\ldots,N_t^1\\
\theta+\alpha N_t^1,& i=N_t^1+1.
\end{array}\right.\end{equation}
After choosing $v_1(t)$, we define $N_t^2$ according to whether or not $v_1(t)$ is a newly observed vertex: if $v_1(t)=N_t^1+1$, then we define $N_t^2=N_t^1+1$; otherwise, we put $N_t^2=N_t^1$.
We then choose $v_2(t)$ as in \eqref{eq:update} with $N_t^1$ replaced by $N_t^2$.
When generating a network with directed edges, we orient edges to point from $v_1(t)$ to $v_2(t)$; in the undirected case, the edge between $v_1(t)$ and $v_2(t)$ has no orientation.
We write $G_n$ to denote the network generated after $n$ steps of this procedure.

\begin{figure}
\includegraphics[scale=0.8]{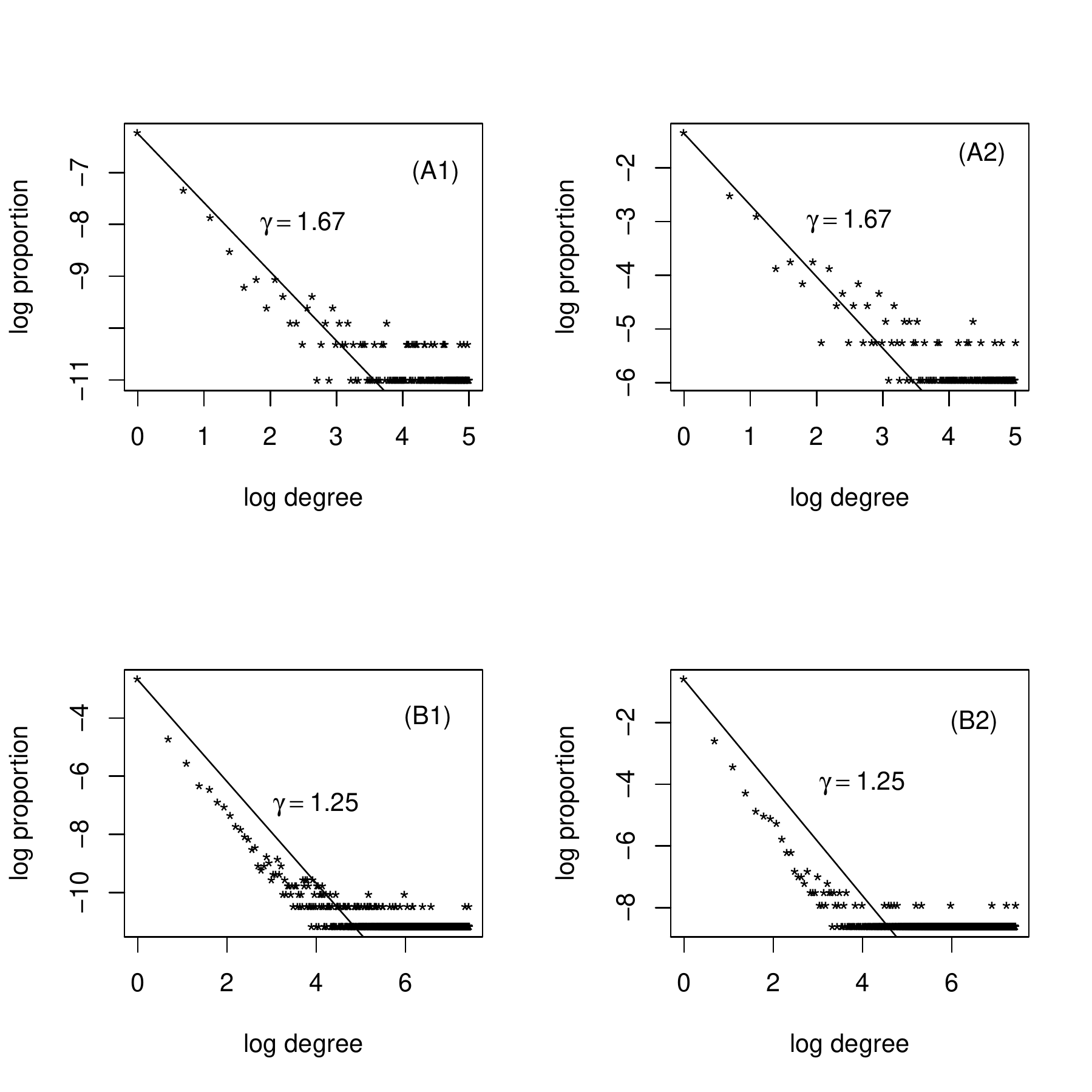}
\caption{Simulation results showing degree distribution of networks and their projection to a simple network by removing multiple edges.  
(A1) Network generated from model with parameters $(\alpha,\theta)=(0.67,1)$, (B1) Network generated from model with parameters $(\alpha,\theta)=(0.25,1)$, (A2) Simple network obtained by reducing multiple edges to single edge in (A1) network, (B2) Simple network obtained by reducing multiple edges to single edge in (B1) network.
Results suggest that the generated network and its induced simple network both exhibit power law of similar degree.}
\label{fig:powerlaw}
\end{figure}

The above generative model produces a sequence of networks $(G_n)_{n=1,2,\ldots}$, where $G_n$ has $n$ edges and a random number of vertices $N_n$.
For $k=1,2,\ldots$, we write $N_n(k)$ to denote the number of vertices in $G_n$ with degree $k$, so that $N_n=\sum_{k\geq1}N_n(k)$.
From properties of the generating mechanism in \eqref{eq:update}\cite{Pitman2005}, the empirical degree distributions $p_n(k)=N_n(k)/N_n$ converge to $\alpha\cdot k^{-(\alpha+1)}/\Gamma(1-\alpha)$, where $\Gamma(t)=\int_{0}^{\infty}x^{t-1}e^{-x}dx$ is the gamma function.
The simulation results in Figure \ref{fig:powerlaw} verify this property of our model.
Moreover, the expected number of vertices satisfies 
\begin{equation}\label{theta}E(N_n)\sim\frac{\Gamma(\theta+1)}{\alpha \cdot \Gamma(\theta+\alpha)}(2n)^\alpha,\quad\text{as }n\rightarrow\infty.\end{equation}
Given an observed power law exponent $1<\gamma<2$, we can use these two properties to estimate the model parameters $\alpha$ and $\theta$ by setting $\alpha=\gamma-1$ and choosing the value of $\theta$ so that Equation \eqref{theta} is satisfied by the observed network.
The estimated parameters in the caption of Figure \ref{fig:alldatasets} were obtained by this method.

Remarkably, we can express the probability distribution of the random network $G_n$ in closed form by:
\begin{equation} \label{ewens}
\pr({G}_n=G)=\alpha^{\#V(G)}\frac{(\theta/\alpha)^{\uparrow\#V(G)}}{\theta^{\uparrow(2n)}}\prod_{v:\,\text{deg}(v)>1}(1-\alpha)^{\uparrow(\text{deg}(v)-1)}
\end{equation}
where $G$ is any network with $n$ edges that can be generated by \eqref{eq:update}, $\text{deg}(v)$ is the degree of vertex $v$ in $G$, $\#V(G)$ is the number of vertices in $G$, and $x^{\uparrow j}=x(x+1)\cdots(x+j-1)$ is the ascending factorial function.
A further important property of ${G}_n$ is that its distribution \eqref{ewens} is independent of the order in which edges arrive during network formation.
As this information is typically unavailable for network data, viable statistical models should be agnostic to it.
Nevertheless, many network models, including preferential attachment models, do depend on the order of arrival, severely limiting the scope of statistical inferences \cite{McCullagh2002}.
Under our model, this lack of information has no adverse consequences.
Therefore, we expect that the discovery of \eqref{ewens} and its intuitive explanation of network formation should lead to significant progress in statistical network analysis.

Our generating mechanism allows for self-loops and multiple edges between vertices, features common in  many of the interaction networks we consider.
For the Enron and actors networks, respectively, self-loops correspond to emailing oneself and acting in the same movies as oneself, while multiple edges reflect an exchange of multiple emails between individuals and a casting of the same actors in multiple movies.
Although these features may be present in the underlying real world phenomenon, network datasets are often simplified by reducing multiple edges to a single edge.
In fact, of the four networks in Figure \ref{fig:alldatasets}, only the actor collaboration network dataset records multiple edges.
Thus, Figure \ref{fig:alldatasets} suggests that atypical scaling is not only present in interaction networks with multiple edges but also in their projection to a simple network by reducing multiple edges to a single edge.
Figure \ref{fig:powerlaw} demonstrates that our model preserves the same scaling under this operation.

The parameters of our model have a clear interpretation in terms of the network generating mechanism.
In \eqref{eq:update}, we see that $\alpha$ controls the rate at which a vertex accumulates edges, leading to the explicit relationship between $\alpha$ and the power law exponent $\gamma=\alpha+1$.
Given the value of $\alpha$, $\theta$ controls the growth of vertices, with large values corresponding to faster growth.
The $\theta$ parameter exhibits its biggest influence at the beginning of network formation.
High estimates of $\theta$ for the Wikipedia and Facebook networks support the conclusion that most votes in Wikipedia elections involve users who did not participate in previous elections and the formation of Facebook social circles begins with rapid addition of new individuals.
The moderate estimate of $\theta$ for the actors network supports the opposite conclusion; indeed, a core of the same movie actors are cast repeatedly while the majority of actors struggle for roles.
The negative estimate of $\theta$ for the Enron network reflects the tendency for communication within a closed group to outpace the rate at which new team members are introduced.

The occurrence of power law exponent between 1 and 2 in several common network datasets brings forth a previously undetected feature of real world network evolution.
While our sequential construction in \eqref{eq:update} and preferential attachment dynamics both grow the network in a size-biased manner---higher degree vertices accumulate edges at a faster rate---network growth under our model is driven by the addition of edges, which accurately reflects the dynamics of the underlying network.
Preferential attachment models, on the other hand, achieve the complementary power law behavior by sequential addition of vertices, behavior not reflective of networks (A)-(D).
Even with state of the art methods \cite{BickelChen2009PNAS,BickelChenLevina2011}, statistical network models are not sufficiently robust to answer many questions of practical interest \cite{McCullagh2002}.
Our model also possesses fundamental statistical properties that lead to straightforward estimation of the parameters $\alpha$ and $\theta$ and, hence, the power law exponent.
Explicit calculation of the distribution \eqref{ewens} opens the door to much more detailed statistical analyses by likelihood-based and Bayesian techniques.
Although power law exponent in this range has not received much attention, we expect that it is widespread in real world interaction networks.
Our framework should lay the foundation for future investigations, both scientific and mathematical.

\end{document}